\documentstyle[11pt,menu99,epsfig]{article}

\begin{document}
    \setlength{\baselineskip}{2.6ex}

\title{Threshold Behaviour of Meson-Nucleon-S${}_{11}$(1535) Vertexfunctions
and Determination of the S${}_{11}$(1535) Mixing Angle}
\author{
F. Kleefeld\thanks{Email: kleefeld@theorie3.physik.uni-erlangen.de / Publication No.
FAU-TP3-99/7} \\
{\em University of Erlangen--N\"urnberg, Institute for Theoretical Physics III, 
Staudtstr. 7,
D-91058 Erlangen, Germany} }

\maketitle

\begin{abstract}
\setlength{\baselineskip}{2.6ex}
A new method for the determination of the spin-1/2-3/2-mixing angle and the range
parameter of the quarkmodel wavefunction of the
resonance $S^-_{11}(1535)$ is presented. The method ist based on a quantitative
calculation of the total cross section of $pp\rightarrow pp\eta$ at threshold. 
The quantitative on-shell treatment of ISI and FSI is discussed. 
\end{abstract}

\setlength{\baselineskip}{2.6ex}

\section*{INTRODUCTION}

Nearly all calculations for the determination of mixing angles 
between different spin-components of baryon wavefunctions are based on 
spectroscopic models. Beside their advantages these models also contain some
problems which are not dissolved yet.\\
The goal of the presented work is an independent way of the determination of 
mixing angles and range parameters in the harmonic oscillator description of
nucleonic resonances within the nonrelativistic quark model. Although the
approach in a first step is applied to the spin--1/2--3/2--mixing angle $\theta$ 
in the negative parity resonance $S_{11}(1535)$ (the underlying nature of this 
resonance is at the moment object to various speculations), the approach also can
be applied to many other resonances.\\
The idea of the presented method is as
follows: In a first step one has to calculate the total
cross section of a process which is dominated by the resonance under
consideration in such a way that the only remaining free parameters of the 
calculation are the mixing angles and the range parameters of the harmonic oscillator
wavefunctions of the resonance and the involved ground state nucleons. In a second step the parameters are obtained by
adjusting the calculated to the experimentally measured cross section.\\
One process dominated mainly by the $S_{11}(1535)$ is the reaction
$pp\rightarrow pp\eta$ at threshold. In order to determine the desired mixing
parameters one first has to get a {\em quantitative} understanding not only of the
{\em short range dynamics} of the process, but also of the {\em initial and final
state interactions} (ISI and FSI) between the in- and outgoing particles. 
Main investigations along this line have been performed in $ $\cite{1} and 
references therein.

\section*{THE REACTION $pp\rightarrow pp\eta$ AT THRESHOLD}

The reaction $pp\rightarrow pp\eta$ at threshold is described within a
relativistic meson-exchange model. The virtual $S_{11}(1535)$ is
excited/deexited by the exchange of $\delta$--, $\sigma$--, $\pi$--, 
$\eta$--, $\rho$-- and $\omega$--mesons, while it is deexcited/excited by the
produced $\eta$.
\subsection*{Total Cross Section}
Using relativistic normalisations for spinors, creation and annihilation
operators it is straight forward to write down the expression for the total 
cross section of $pp\rightarrow pp\eta$ ($P_i=p_1+p_2$,
$s=P_i^2$, ${\cal F} (s)=2 \sqrt{\lambda (s,m^2_p, m^2_p)}$, 
$\lambda (x,y,z)= x^2 + y^2 + z^2 - 2 (xy+yz+zx)$):
\begin{eqnarray}
\lefteqn{ \sigma_{pp\rightarrow pp\eta}(s) \quad =} \nonumber \\
 & = & \frac{1}{2!} \frac {1}{{\cal F} (s)} \frac{1}{(2\pi)^5}
 \int \! \frac{d^{\,3} p_{1^\prime}}{
2 \, \omega_{p} (|\vec{p}_{1^\prime}|)} \;
\frac{d^{\,3} p_{2^\prime}}{
2 \, \omega_{p} (|\vec{p}_{2^\prime}|)} \;
\frac{d^{\,3} p_{3^\prime}}{
2 \, \omega_{\,\eta} (|\vec{p}_{3^\prime}|)} \;
\;
 \delta^4 (p_{1^\prime} + p_{2^\prime} + p_{3^\prime} - P_i\,) 
\, \overline{\;\left|T_{fi} \right|^2} \nonumber \\
\end{eqnarray}
The combinatorial factor of $1/2\,!$ is due to the two outgoing identical
protons. For the relativistic description of $pp\rightarrow pp\eta$ the following
5 independent invariants are chosen: 
\begin{equation} s = (p_1+p_2)^2 \;,\; s_1 = (p_{1^\prime}+p_{2^\prime})^2
\;,\; s_2 = (p_{2^\prime}+p_{3^\prime})^2 \;,\; t_1 = (p_1 - p_{1^\prime})^2 \;,\; t_2 = (p_2 - p_{3^\prime})^2
\end{equation}

\subsection*{(On--Shell) Watson-Migdal-Approach to ISI and FSI}
In order to get a {\em quantitative} understanding of the ISI and FSI it is
crucial to recall the steps which lead to that what the people nowadays call``Watson-Migdal approximation" and what is wrongly implemented in nearly all
threshold meson production calculations by now: 
The system of colliding and produced particles is described by an overall 
Hamilton operator $H = K + V = K + W + U$, while $K$ is the kinetic energy
operator, $W$ is the particle number nonconserving short range interaction
potential and $U$ is the particle number conserving interaction potential of 
long range. For convenience additionally one can define the long range Hamilton operator
$h=K+U$. The corresponding eigenstates to the operators $H$,
$h$ and $K$ fulfil the following Schr\"odinger equations:
\begin{equation}
(E-H)|\psi^\pm_\alpha > = 0 \quad , \quad
(E-h)|\chi^\pm_\alpha > = 0 \quad , \quad
(E_n-K)|\varphi^\pm_n > = 0 
\end{equation}
The corresponding Lippmann--Schwinger equations are:
\begin{eqnarray}
|\psi^\pm_\alpha > \quad = & \displaystyle
|\varphi_\alpha > + \frac{1}{E-H\pm i\varepsilon} \; V \, |\varphi_\alpha >
& = \quad |\chi^\pm_\alpha > + \frac{1}{E-H\pm i\varepsilon} \; W \, |\chi^\pm_\alpha > \nonumber \\
|\chi^\pm_\alpha > \quad = & \displaystyle
|\varphi_\alpha > + \frac{1}{E-h\pm i\varepsilon} \; U \, |\varphi_\alpha >
& = \quad |\varphi_\alpha > + \frac{1}{E-K\pm i\varepsilon} \; T^\pm_{el}(E,K) \, |\varphi_\alpha
> \quad
\end{eqnarray}
Defining the free propagator $G_0^\pm (E,K)= (E-K\pm i\varepsilon)^{-1}$ and inserting complete sets of
free asymptotic states the T-matrix $T_{fi}$ in Eq. (1) can be
separated in a short and long ranged part via 
($T_{\beta\alpha} \; = \; 
<\chi^-_\beta|W|\psi^+_\alpha > \; = \; <\psi^-_\beta|W|\chi^+_\alpha >$):
\begin{eqnarray}
T_{\beta\alpha} \quad \simeq \quad
<\chi^-_\beta|W|\chi^+_\alpha > & = &
 <\varphi_\beta|(1+\sum\limits_n T^+_{el}(E,K)|\varphi_n > G_0^+ (E,E_n) <\varphi_n |)W 
\nonumber \\
 & & ( 1 + \sum\limits_m |\varphi_m > G_0^+ (E,E_m) <\varphi_m |T^+_{el}(E,K))|\varphi_\alpha>
\end{eqnarray}
As the in- and outgoing particles in $pp\rightarrow pp\eta$ are nearly on-shell one can approximate the 
free propagator $G_0^\pm (E,K)$ in the following way: 
\begin{equation}
G_0^\pm (E,K) \; = \; \frac{1}{E-K\pm i\varepsilon} \; = \; 
P \,\frac{1}{E-K} \; \mp \; i\pi \delta(E-K) \; \approx \;
\mp \; i\pi \delta(E-K)
\end{equation}
At this point it is useful to introduce the following relative lab wavenumbers for each pair of
particles in the initial and final state ($s_3 = (p_{3^\prime}+p_{1^\prime})^2 = s-s_1-s_2+2m^2_p+m^2_\eta$):
\begin{equation}
k = \frac{\sqrt{\lambda (s,m^2_p,m^2_p)}}{2(m_p+m_p)} \; , \;
\kappa = \frac{\sqrt{\lambda (s_1,m^2_p,m^2_p)}}{2(m_p+m_p)} \; , \;
\kappa_1 = \frac{\sqrt{\lambda (s_3,m^2_p,m^2_\eta)}}{2(m_p+m_\eta)} \; , \;
\kappa_2 = \frac{\sqrt{\lambda (s_2,m^2_p,m^2_\eta)}}{2(m_p+m_\eta)} 
\end{equation} 
and the corresponding wavevectors $\vec{k}$, $\vec{\kappa}$, $\vec{\kappa}_1$, $\vec{\kappa}_2$.
Consider now the case, when the only two particles taking part in the ISI and FSI are the two in-- and
outgoing protons. In this case the subsystem of the two protons  
interacting via the long range potental $U$ with phaseshifts
$\delta_\ell $ ($\ell$ = orbital angular momentum) can be treated within
the framework of standard nonrelativistic scattering theory. 
By application of the on-shell approximation of Eq. (6) the T-matrix in Eq. (5) can be 
transformed to ($\vec{k}^{\,\prime} = \vec{\kappa}$, 
$k^{\,\prime} = \kappa$): 
\begin{eqnarray} T_{fi} \, \simeq \, T(\vec{k}^{\,\prime},\ell^{\,\prime};\vec{k},\ell) & = &
\left( 1+ i e^{i\delta_{\ell^\prime}(k^\prime)}\sin\delta_{\ell^\prime}(k^\prime) \right) 
<\vec{k}^{\,\prime},\ell^{\,\prime}|W|\vec{k},\ell>
\left( 1+ i e^{i\delta_{\ell}(k)}\sin\delta_{\ell}(k)\right) \nonumber \\
 & = &
\frac{k^\prime \cot\delta_{\ell^\prime}(k^\prime)}{k^\prime ( \cot\delta_{\ell^\prime}(k^\prime)-i) } 
<\vec{k}^{\,\prime},\ell^{\,\prime}|W|\vec{k},\ell>
\frac{k \cot\delta_{\ell}(k)}{k (\cot\delta_{\ell}(k)-i)} \nonumber \\
 & = &
\frac{\mbox{Re} f_{\ell^\prime}(k^\prime)}{f_{\ell^\prime}(k^\prime) } 
<\vec{k}^{\,\prime},\ell^{\,\prime}|W|\vec{k},\ell>
\frac{\mbox{Re} f_{\ell}(k)}{f_{\ell}(k) } \; = \; T(\mbox{FSI}) \;
{\bar{T}}_{fi} \; T(\mbox{ISI})
\end{eqnarray}
Here the Jost functions $f_\ell(k)$ were introduced by 
$e^{i\delta_{\ell}(k)}\sin\delta_{\ell}(k)=-(\mbox{Im}f_{\ell}(k))/f_{\ell}(k)$. It is worth to
mention that the ISI-- and FSI--factors of Eq. (8) have the correct limit for vanishing ISI or FSI.
If there is no ISI or FSI, the phaseshifts will vanish and the corresponding factors will go to 1. 
The effect of the Watson-Migdal approach is a factorization of
$T_{fi}$ into a short ranged T-matrix ${\bar{T}}_{fi}$ and ISI- and
FSI-factors $T(\mbox{ISI})$ and $T(\mbox{FSI})$.
\newpage
The the factor $T(\mbox{ISI})$ is only a function of $s$, while the short ranged
T-matrix ${\bar{T}}_{fi}$ close to threshold shows up to be a slowly varying 
function of the phasespace integration variables and therefore can be set to its
threshold value ${\bar{T}}^{\,\scriptsize \mbox{thr}}_{fi}$. Using this
observation both factors can be drawn in front of the phasespace integral in Eq.
(1) with the result:
\begin{eqnarray}
\lefteqn{ \sigma_{pp\rightarrow pp\eta}(s) \quad \simeq } \nonumber \\
 & \simeq & 
\frac{\left|{\bar{T}}^{\,\scriptsize \mbox{thr}}_{fi} 
 \; T(\mbox{ISI})\right|^2}{2! {\cal F} (s) (2\pi)^5}
  \int \! \frac{d^{\,3} p_{1^\prime}}{
2 \, \omega_{p} (|\vec{p}_{1^\prime}|)} \;
\frac{d^{\,3} p_{2^\prime}}{
2 \, \omega_{p} (|\vec{p}_{2^\prime}|)} \;
\frac{d^{\,3} p_{3^\prime}}{
2 \, \omega_{\,\eta} (|\vec{p}_{3^\prime}|)} \;
\;
 \delta^4 (p_{1^\prime} + p_{2^\prime} + p_{3^\prime} - P_i\,) 
\, \left| T(\mbox{FSI}) \right|^2 \nonumber \\
 & = &
\frac{\left|{\bar{T}}^{\,\scriptsize \mbox{thr}}_{fi} 
 \; T(\mbox{ISI})\right|^2}{2! {\cal F} (s) (2\pi)^5}
\; R^{\scriptsize FSI}_3(s)
\end{eqnarray}
In the last line of Eq. (9) the final state interaction modified phasespace
integral $R^{\scriptsize FSI}_3(s)$ for the three body final state $pp\eta$ 
has been defined. To expand the energy dependence of the total cross section at
threshold one usually expresses the cross section in terms of the dimesionless
variable $\eta = \sqrt{\lambda(s,m^2_\eta,s^{\scriptsize \mbox{min}}_1)/(4 s \, m^2_\eta)}$
instead of the square of the cm-energy $s$. The quantity $\eta$ being the maximum momentum of the
produced $\eta$-meson in the cm-frame vanishes at threshold. It is easy to 
derive $s^{\scriptsize \mbox{min}}_1=(2m_p)^2$. Introducing $T_{\scriptsize
\mbox{lab}}=(s-4m^2_p)/(2m_p)$ and $\mu = m_p/m_\eta$ the phaseshifts of the
long range potential between the incoming two protons can be expanded at threshold:
\begin{equation} \delta (T_{\scriptsize \mbox{lab}}) = 
\underbrace{\delta (T^{\,\scriptsize \mbox{thr}}_{\scriptsize \mbox{lab}})}_{\displaystyle =:\,\delta^{(0)}} + 
(T_{\scriptsize \mbox{lab}}
-T^{\,\scriptsize \mbox{thr}}_{\scriptsize \mbox{lab}}) 
\delta^\prime (T^{\,\scriptsize \mbox{thr}}_{\scriptsize \mbox{lab}}) + \ldots =
\delta^{(0)} + 
\underbrace{m_\eta \left(1+\frac{1}{2\mu}\right)^2
\delta^\prime (T^{\,\scriptsize \mbox{thr}}_{\scriptsize \mbox{lab}})}_{\displaystyle 
=:\,\delta^{(2)}}\eta^2 + O(\eta^4)
\end{equation}  
In terms of these expansion coefficents the factor $T(\mbox{ISI})$ can be
expanded at threshold:
\begin{equation}
T(\mbox{ISI}) \; = \; 1+ i \, e^{i\delta(k)}\sin\delta(k) \; = \;
1+ i e^{i\delta^{(0)}}\sin\delta^{(0)} + i \, e^{2i\delta^{(0)}}\delta^{(2)} \eta^2 +
O(\eta^4) 
\end{equation}
At the moment there is no clear knowledge of the values of $\delta^{(0)}$ and
$\delta^{(2)}$ at the production threshold of $pp\rightarrow pp\eta$, but as the
incoming relative proton momentum is very large the interaction time is very
short, so that one might assume $\delta^{(0)}$ and $\delta^{(2)}$ to be zero
which yields $T(\mbox{ISI})\approx 1$, while the VPI-values 
$\delta^{(0)}\approx -60^o$, $\delta^{(2)}\approx 0$ lead to 
$|T(\mbox{ISI})|\approx 1/2$.\\
For the complete determination of $T(\mbox{FSI})$ one obviously has to solve the
Faddeev equations for the outgoing $pp\eta$--system. As this is at the moment out
of the scope of this work only the leading terms of the Fadeev expansion are taken
into account:
\begin{equation} T(\mbox{FSI}) \; \approx \;
\frac{\kappa \cot\delta_{1^\prime 2^\prime}(\kappa)}{
\kappa (\cot\delta_{1^\prime 2^\prime} (\kappa)-i)}
\; + \; \frac{\kappa_1 \cot\delta_{3^\prime 1^\prime}(\kappa_1)}{
\kappa_1 (\cot\delta_{3^\prime 1^\prime} (\kappa_1)-i)}
\; + \; \frac{\kappa_2 \cot\delta_{2^\prime 3^\prime}(\kappa_2)}{
\kappa_2 (\cot\delta_{2^\prime 3^\prime} (\kappa_2)-i)}
\; - \; 2 
\end{equation}
The various terms in Eq. (12) which denote the FSI between each pair of particles
in the final state can be expanded by effective range expansions. For the
outgoing pp--system the s-wave nuclear effective range expansion is
given by:
\begin{equation} \kappa \cot\delta_{1^\prime 2^\prime} (\kappa) = - \frac{1}{a} + \frac{r}{2}
\kappa^2 + O(\kappa^4) \quad \mbox{with} \quad 
a\approx -17.1\; \mbox{fm} \; , \; r\approx 0 \ldots 2.84\; \mbox{fm} 
\end{equation}
Taking into account only pp-FSI using the shape independent effective
range expansion Eq. (13) the FSI-modified phasespace integral can be
expanded at threshold ($\bar{a}=a\, m_\eta$):
\begin{eqnarray} R^{\scriptsize FSI}_3(s) & \simeq & 
  \int \! \frac{d^{\,3} p_{1^\prime}}{
2 \, \omega_{p} (|\vec{p}_{1^\prime}|)} \;
\frac{d^{\,3} p_{2^\prime}}{
2 \, \omega_{p} (|\vec{p}_{2^\prime}|)} \;
\frac{d^{\,3} p_{3^\prime}}{
2 \, \omega_{\,\eta} (|\vec{p}_{3^\prime}|)} \;
\;
 \delta^4 (P_f - P_i\,) 
\, \left| \frac{\displaystyle - \frac{1}{a} + \frac{r}{2} \kappa^2}{
\displaystyle - \frac{1}{a} + \frac{r}{2} \kappa^2 -i\kappa} \right|^2 \nonumber
\\
 & = & \frac{\pi^3}{4} \; m_\eta^2 \; \frac{\sqrt{1+2\mu}}{2\mu} 
 \left\{ \frac{\eta^4}{2^2} 
 - \frac{\eta^6}{2^6} [4 (2\mu)^{-2} + 7 + 2 \bar{a}^2 + 2\bar{a}^2 (2\mu)] 
 + O (\eta^8) \right\}
\end{eqnarray}
It is interesting to observe that even the $\eta^6$-term is independent of the
effective range $r$. \newpage 
\subsection*{Relativistic Meson-Exchange-Amplitudes}
For the relativistic meson-exchange model the following compact expression for 
the ${}^3P_0\rightarrow {}^1S_0s$ threshold transition amplitude ${\bar{T}}^{\,\scriptsize
\mbox{thr}}_{fi}$ of $pp\rightarrow pp\eta$ is obtained $ $\cite{1,2}
($m_N \simeq m_p$):
\begin{eqnarray}
{\bar{T}}^{\,\scriptsize \mbox{thr}}_{fi}
& = & 2\, m_{\scriptscriptstyle N} \,
\sqrt{m_\eta \; (m_\eta + 4\, m_{\scriptscriptstyle N} \,)} \,
 \; [ \,
(\, X_{\,\delta } + X_{\,\sigma } \, ) \, (m_{\scriptscriptstyle N} + m_\eta ) - 
(\, X_{\,\pi } + X_{\,\eta } \, ) \, (m_{\scriptscriptstyle N} - m_\eta ) \, +
 \nonumber \\
 & + & 
(\, Y_{\,\delta } + Y_{\,\sigma } - Y_{\,\pi } - Y_{\,\eta } \, ) \, (m_{\scriptscriptstyle N} + m_\eta ) + 
 M_\delta + M_\sigma - M_\pi - M_\eta
 \, ]
 \, - \nonumber \\
 & - &  X_\rho \; m_{\scriptscriptstyle N} \;
 [ \,
 4 \, (m_\eta - 2\, m_{\scriptscriptstyle N}) +
 K_\rho \, (5 m_\eta - 4\, m_{\scriptscriptstyle N})
 \, ] \; + \nonumber \\
 & + & 
 [ \,
Y_{\,\rho } \, (m_{\scriptscriptstyle N} + m_\eta ) + 
 \tilde{M}_\rho 
 \, ]
 \; [\, K_{\rho} 
 \; (m_\eta - 4\, m_{\scriptscriptstyle N})
 - 8\, m_{\scriptscriptstyle N} \, ]  - \nonumber \\
 & - & X_\omega \; m_{\scriptscriptstyle N} \;
 \left[ \,
 4 \, (m_\eta - 2\, m_{\scriptscriptstyle N}) +
 K_\omega \, (5 m_\eta - 4\, m_{\scriptscriptstyle N})
 \, \right] \; + \nonumber \\
 & + & 
 [ \,
Y_{\,\omega } \, (m_{\scriptscriptstyle N} + m_\eta ) + 
 \tilde{M}_\omega 
 \, ]
 \; [\, K_{\omega} 
 \; (m_\eta - 4\, m_{\scriptscriptstyle N})
 - 8\, m_{\scriptscriptstyle N} \, ]  \qquad (K_\rho \approx 6.1, K_\omega \approx 0)
\end{eqnarray}
Here I used the following abbreviations ($\phi\in \{\delta,\sigma,\pi,\eta,\rho,\omega\}$)
($M_{S_{11}}:=m_{S_{11}}- i \,\Gamma_{S_{11}}/2$) ($D_\phi (q^2):=(q^2-m_\phi^2)^{-1}$)
($q^2:=- m_p m_\eta$, $p^2:=m_p \; (m_p-2 m_\eta)$,
$P^2:=(m_p+m_\eta)^2$):
\begin{eqnarray}
 & & X_{\,\phi }
 :=  
 D_\phi (q^2) \; g_{\, \phi NN} (q^2) \;
 D_{S_{11}} (p^2) \; g^{\, \ast}_{\phi NS_{11}} (q^2) \;  
 g_{\eta NS_{11}} (m^2_{\, \eta}) \nonumber \\
 & & Y_{\,\phi }
 :=  
 D_\phi (q^2) \; g_{\, \phi NN} (q^2) \;
 D^R_{S_{11}} (P^2) 
 \;  
g^{\, \ast}_{\eta NS_{11}^R} (m^2_\eta) \;  
 g_{\phi NS_{11}^L} (q^2 ) \nonumber \\
 & & M_\phi := X_\phi \, m_{S_{11}} + Y_\phi \, M_{S_{11}} \quad , \quad
 \tilde{M}_\phi := - X_\phi \, m_{S_{11}} + Y_\phi \, M_{S_{11}} \nonumber \\
 & & 
D^R_{S_{11}} (P^2):=(P^2-M^2_{S_{11}})^{-1} \; , \quad
D_{S_{11}} (p^2):=(p^2-m^2_{S_{11}})^{-1} 
\end{eqnarray}
\subsection*{Coupling Constants and Vertexfunctions}
The meson--nucleon--$S_{11}(1535)$ couplings are in general complex and show a
strong nontrivial momentum sensitivity. For that reason it is not enough to 
evaluate Eq. (16) applying the commonly used on-shell coupling constants combined 
with standard monopole or dipole formfactors. Hence a model has been
developed to estimate the real and imaginary parts of the couplings $ $\cite{1,3}.
The real parts of the couplings are derived within the framework of a 
nonrelativistic quark model, in which the ground state nucleon and the 
$S_{11}(1535)$ wavefunctions are described by harmonic oscillator solutions. The
radial wavefunctions $R_p$, $R^{(23)}_{S_{11}}$ and $R^{(1,23)}_{S_{11}}$ of the ground state nucleon and the 
$S_{11}(1535)$ are determined by one unique range parameter $b$ (Here $\rho_{23}$, 
$\rho_{1,23}$ are 3--quark--Jacobian coordinates!):
\begin{equation} R_p \propto e^{-b^2 (\rho^2_{23} + \rho^2_{1,23})/2} , \; 
R^{(23)}_{S_{11}} \propto b \,\rho_{23} \, e^{-b^2 (\rho^2_{23} + \rho^2_{1,23})/2}, \; 
R^{(1,23)}_{S_{11}} \propto b \,\rho_{1,23} \, e^{-b^2 (\rho^2_{23} + \rho^2_{1,23})/2}
\end{equation}
The imaginary parts of the couplings are calculated close the threshold from
relevant lowest order meson loop corrections to the bare couplings. 
\section*{FIRST RESULTS}
Taking into account only $\pi$--, $\eta$--, $\rho$-- and $\omega$--exchange we
observe by reproducing the experimental total cross section of $pp\rightarrow
pp\eta$ that the following mixing
parameters are favourable: $\theta \approx - 5^o$, $b^{-1}\approx 0.5$ fm. Because
of the uncertainties in the bare meson-nucleon-$S_{11}(1535)$ couplings and the
still improvable treatment of ISI and FSI these numbers are by now preliminary. 
The model gives for the first time a quantitative prediction of the relative
phases between the $\pi$--, $\eta$--, $\rho$-- and $\omega$--exchange amplitudes.
An extension of the effective range formalism for ISI and FSI to
Coulomb-interactions is on the way.
\bibliographystyle{unsrt}

\end{document}